\documentclass[aps,superscriptaddress,eqsecnum,nofootinbib,showpacs,preprintnumbers]{revtex4}
\usepackage{graphicx,epsfig}
\usepackage{amsmath}
\usepackage {amssymb}
\usepackage{pstricks}

\def\be{\begin{equation}}
 \def\ee{\end{equation}}
 \def\bea{\begin{eqnarray}}
 \def\eea{\end{eqnarray}}

\makeindex

\begin{document}

\title{Motion of magnetically charged particles in a magnetically charged stringy black hole spacetime}
\author{P. A. Gonz\'{a}lez}
\email{pablo.gonzalez@udp.cl}
 \affiliation{Facultad de
Ingenier\'{i}a y Ciencias, Universidad Diego Portales, Avenida Ej\'{e}rcito
Libertador 441, Casilla 298-V, Santiago, Chile.}
\author{ Marco Olivares }
\email{marco.olivaresr@mail.udp.cl}
\affiliation{ Facultad de Ingenier\'ia y Ciencias, Universidad Diego Portales,
Avenida Ej\'ercito Libertador 441, Casilla 298-V, Santiago, Chile.}
\author{Eleftherios Papantonopoulos}
\email{lpapa@central.ntua.gr}
\affiliation{Department of
Physics, National Technical University of Athens, Zografou Campus
GR 157 73, Athens, Greece.}
\author{Joel Saavedra}
\email{joel.saavedra@ucv.cl}
\affiliation{Instituto de
F\'{i}sica, Pontificia Universidad Cat\'olica de Valpara\'{i}so,
Casilla 4950, Valpara\'{i}so, Chile.}
\author{Yerko V\'{a}squez}
\email{yvasquez@userena.cl}
\affiliation{Departamento de F\'{\i}sica y Astronom\'ia, Facultad de Ciencias, Universidad de La Serena,\\
Avenida Cisternas 1200, La Serena, Chile.}
\date{\today}

\vspace{0.5cm}

\begin{abstract}
{We study the motion of massive particles with electric and magnetic charges in the background of a magnetically charged Garfinkle-Horowitz-Strominger stringy black hole. We solve analytically the equations of motion of the test particles and we describe the orbital motion in terms of the Weierstrass elliptic functions. We find that there are critical values of the magnetic charge of the black hole and the magnetic charge of the test particle which characterize the  bound and unbound orbits and we study
two observables,
the perihelion shift
 and the Lense-Thirring effect.
The trajectories depend on the electric and magnetic charges of the test particle. While the angular-motion depends on the electric charge of the test particle, the $r$ and $t$-motion depends on the mass and the magnetic charge of the test particle.}
\end{abstract}
\keywords{Black Holes;  Elliptic Functions; String Theory}
\maketitle 
\tableofcontents

\newpage
\section{Introduction}

The effective four-dimensional field theory of a heterotic string theory has local black hole solutions
which they can have properties very different from those
that appear in the black hole solutions of general relativity (GR).
A static stringy charged black hole solution of this theory
was found in \cite{Gibbons:1987ps}, and independently in
 \cite{Garfinkle:1990qj}, know as GHS black hole.
Thereafter, much research has  been performed in the
context of the heterotic string theory. In \cite{Sen:1992ua}
 a black hole solution in four dimensions carrying
mass, charge and angular momentum was found and also
the extremal limit of this black hole solution was discussed. Additionally, this work was further extended in \cite{Sen:1994eb} and
a general electrically charged, rotating black hole solution in the heterotic string theory compactified on a six-dimensional torus was found.
Then it was shown in \cite{Hassan:1991mq}, that given a classical solution of the heterotic string theory other
 classical solutions can be generated by transforming the original solution. Using this method black string solutions in six dimensions were constructed carrying electric charge, and both,
electric and magnetic type antisymmetric tensor gauge field charge. On the other hand, the introduction of the basic aspects of  solitons and
black holes and duality in string theory  \cite{Khuri:1997gw}, stimulated the search of static solutions
of electrically and magnetically charged dilaton black holes with  various  topologies and with the introduction of a  cosmological constant \cite{Gao:2005xv}.

The properties of  charged black holes in string theory can be revealed   studying the geodesics around these solutions. This is because except the information we get solving the classical
equations of motion in the form of Einstein equations we also get information about stringy corrections due to the string coupling which is of the order of Planck scale.
The study of null geodesics in the electrically charged GHS black hole was carried out in Refs. \cite{Fernando:2011ki, Soroushfar:2016yea}, and the timelike geodesics were analyzed in \cite{Maki:1992up, Olivares:2013jza, Blaga:2014lva, Blaga:2014spa, Soroushfar:2016yea}.
Additionally, in \cite{Villanueva:2015kua} the gravitational Rutherford scattering and Keplerian orbits were studied in   the GHS black hole background.

Magnetically charged black holes have also been studied and their stability was analysed in connection to strong coupling. In the Maxwell-Einstein theory generalized magnetically charged Reissner-Nordstr\"om black hole solutions can be found and also in string theory as generalization of electrically charged GHS black hole solutions  \cite{Gibbons:1987ps,Garfinkle:1990qj}. An interesting feature of magnetically charged black holes is their connection to magnetic monopoles. In \cite{Lee:1991qs} it was shown that a magnetically charged Reissner-Nordstr\"om solution develops a classical instability which may lead to a nonsingular magnetic monopole. The magnetic monopoles are hypothetical particles that have not been observed in nature, however, grand unified theories and string theory predict their existence \cite{Wen:1985qj}.
  Also, Dirac showed that the existence of a magnetic monopole in the Universe implies the quantization of the electric charge \cite{Dirac:1931kp}. Additionally, magnetic monopoles appear as regular solutions of the $SU(2)$ Yang-Mills-Higgs theory \cite{tHooft:1974kcl,Polyakov:1974ek}. A magnetic monopole generated by a gravitational field was discussed in \cite{Barriola:1989hx}.

An important issue in gravitational physics is to know if test particles outside the event horizon of a black hole follow stable circular orbits or not.
This information can be provided by studying the geodesics around these black holes. The geodesics of the magnetically charged GHS stringy black hole were analyzed in \cite{Kuniyal:2015uta, Soroushfar:2016yea}, and
it was found 
that there exists no stable circular orbits outside the event horizon of this
stringy black hole for massless test particles. Additionally,  the behaviour of null geodesics has  been used to calculate the absorption cross section for massless scalar waves at high frequency limit or geometric optic limit, because at high frequency limit the absorption cross section can be approximated by the geometrical cross section of the black hole photon sphere $\sigma\approx\sigma_{geo}=\pi b_{c}^2$, where $b_{c}$ is the impact parameter of the unstable circular orbit of photons. Moreover, in \cite{Decanini:2011xi, Decanini:2011xw} this approximation was improved at the high frequency limit by $\sigma \approx \sigma_{geo}+\sigma_{osc}$, where $\sigma_{osc}$ is a correction involving the geometric characteristics of the null unstable geodesics lying on the photon sphere, such as the orbital period and Lyapunov exponent. This approximation was used recently in \cite{Fernando:2017qrd} to evaluate the absorption cross section of electromagnetic waves at high frequency limit.

In this work we will investigate the timelike geodesics around the magnetically charged GHS black hole. We will show that the geodesic structure for this spacetime will reveal that  the motion of charged particles in this   magnetically charged space-time will be very different compared to the motion of photons, giving us important information about the structure and properties of this spacetime. For the photons the motion can be studied only in the equatorial plane due to the fact that photons do not carry electric and magnetic charge, therefore they do not feel the Lorentz force. However, in the case of massive particles with electric and magnetic charges because the  Lorentz force is perpendicular to the particle four-velocity the motion of the charged particles can not be restricted only on a plane. We will also show in this study,  that stringy effects play a crucial role in the behaviour of the orbit, because of the relation of magnetic charge to  the string coupling.

We organize  the work as follows. In Section \ref{HJEOM} after reviewing in brief the magnetically charged GHS spacetime we
present the procedure to obtain the equations of motion of
massive particles in the magnetically charged GHS black hole background. In Section \ref{eqofmot} we analyze and solve the equations of motion   in terms of the Weierstrass elliptic functions. In Section \ref{orbits} we analyze the orbital motion  for a choice  of the parameters. In Section \ref{observables}  we present in  briefly the observables of the perihelion shift and the  Lense-Thirring effect. Finally,
in Section \ref{summ} we summarize our results and discuss possible extensions.

\section{Equations of motion for magnetically charged particles in a magnetically charged stringy black hole}
\label{HJEOM}

In this section we briefly review the magnetically charged black hole in the GHS spacetime (for a review see \cite{Horowitz:1992jp}) and then we present the procedure of obtaining the equations of motion in this spacetime.

The most general action of low energy  heterotic  string theory is given by
\be
\mathcal{S}= \int d^D x \sqrt{-g}\ e^{-2\varphi} \Bigl[ \Lambda +R+4(\nabla \varphi)^2 -
F_{\mu\nu}F^{\mu\nu} - \frac{1}{12} H_{\mu\nu\rho}
H^{\mu\nu\rho}\Bigr]~, \label{action} \ee
where the scalar field $\varphi$ is the dilaton field, $F_{\mu\nu}$ is a Maxwell field,
and the three form $H_{\mu\nu\rho}$ is  related to
a two-form potential  $B_{\mu\nu}$  and the gauge field $A_\mu$
by $H = dB - A \wedge F$
so that $dH = - F\wedge F $. Note that in this action the term $e^\varphi$
plays the role of a coupling constant giving the strength of the stringy effects.

If we set $H$ to zero and make  the conformal transformation of the metric
to rescale $g_{\mu\nu}$ by
$e^{-2\varphi}$ to get a metric with the standard Einstein action \be g^E_{\mu\nu} =
e^{-2\varphi} g_{\mu\nu}~. \ee  The action now becomes (with $\Lambda=0$)
\be
\mathcal{S}=\int d^4x\ \sqrt{-g_E}\ \left(R_E - 2(\nabla\varphi)^2 - e^{-2\varphi}
F^2\right)~. \label{simaction}
\ee

The magnetically charged black hole can be obtained from the
electrically charged solution \cite{Garfinkle:1990qj} by an electromagnetic duality transformation.
From (\ref{simaction}), the equation of
motion for $F_{\mu\nu}$ is
\be
\nabla_\mu\left(e^{-2\varphi} F^{\mu\nu}\right) = 0~. \ee
This implies $\tilde F_{\mu\nu} \equiv e^{-2\varphi} \frac{1}{2}
{\epsilon_{\mu\nu}}^{\rho\sigma}F_{\rho\sigma}$ is curl free.
 The
equations of motion resulting from the action (\ref{simaction}) are invariant under $F \rightarrow \tilde F$,
$\varphi\rightarrow -\varphi$, and $g_E\rightarrow g_E$ and the metric of the magnetically charged black hole is given by
\begin{equation}\label{e1}
 ds^2= -\frac{f(r)}{a(r)}dt^2+\frac{dr^2}{f(r)a(r)}+r^2d\theta^2+ r^2\sin^2\theta\,d\phi^2~,
\end{equation}
where
$$
f(r)\equiv1-{2\,M\over r}~, \quad a(r)\equiv1-{Q^2\over M\, r}~,
$$
and the coordinates are defined in the ranges
$0<r<\infty$, $-\infty<t<\infty$, $0\leq \theta <\pi$, and
$0\leq \phi <2\pi$. $M$ is related to the mass of the spherical object and $Q$ is its magnetic charge.
Note that since $\varphi$ changes sign compared to electrically charged solution, the string coupling
becomes  strong near the singularity for these black holes.

With the aim to study the motion of charged particles around the magnetically charged black hole,  first we derive the equations of motion following the standard approach  \cite{chandra}.  Then, we consider the motion of test particles with mass $m$, electric charge $q$ and magnetic charge $g$ using  the Hamilton-Jacobi formalism. In the Appendix we give a detailed account  of how the Hamilton-Jacobi formalism is connected with the Euler-Lagrange formalism. The Hamilton-Jacobi equation  for the geometry described by metric $g_{\mu\nu}$ for a magnetically and electrically charged test particle is given by
\begin{equation}
2\frac{\partial S}{\partial\tau}=
g^{\mu \nu}\left(\frac{\partial S}{\partial
x^{\mu}}-qA_{\mu}+ig\check{A}_{\mu}\right)\left(\frac{\partial S}{\partial
x^{\nu}}-qA_{\nu}+ig\check{A}_{\nu}\right)~.\label{i.3}
\end{equation}
The field strength $F_{\mu\nu}= A_{\nu,\mu}-A_{\mu,\nu}$ and the dual field strength
$\check{F}_{\mu\nu}=\check{ A}_{\nu,\mu}-\check{A}_{\mu,\nu}$ of the electromagnetic
field are induced by the non-vanishing components of the vector potentials
$A_\mu$ and$\check{A}_{\mu}$
\begin{equation}
A_\phi=-Q\cos\theta~, \quad \check{A}_t=-{iQ\over r}\label{i.4}~.
\end{equation}
The dual field strength is defined by the antisymmetric Levi-Civita symbol
$\varepsilon^{\mu\nu\sigma\tau}$ as
$ \check{F}^{\mu\nu}=e^{-2 \varphi} {i\over 2\sqrt{g^d}}\varepsilon^{\mu\nu\sigma\tau}F_{\sigma\tau}$
with $g^d =-\det||g_{\mu\nu} ||$.
Taking into account the symmetries of the metric under consideration we solve the Hamilton-Jacobi equation using the following ansatz
\begin{equation}
S=-{1\over2}m^2\tau-E\, t+S_{r}(r)+S_{\theta}(\theta)+L\,\phi~, \label{i.5}
\end{equation}
where $E$ and $L$ are identified as the energy and angular momentum
of the test particle. Then, using this ansatz,  Eq. (\ref{i.3}) reads as follows
\begin{equation}
-m^{2}=
-\frac{a(r)}{f(r)}\left[-E+
{g\,Q\over r} \right]^{2}
+f(r)\,a(r)
\left(\frac{\partial S_{r}}{\partial
r}\right)^{2}+{1\over r^{2}}\left(\frac{\partial S_{\theta}}{\partial
\theta}\right)^{2}+{csc^{2}\theta(L+qQ\cos\theta)^{2}\over r^{2}}~.\label{i.6}
\end{equation}
We can obtain  the following radial equation
\begin{equation}
-m^{2}=
-\frac{a(r)}{f(r)}\left[E-
{g\,Q\over r} \right]^{2}
+f(r)\,a(r)
\left(\frac{\partial S_{r}}{\partial
r}\right)^{2}+{k\over r^{2}}~,\label{i.8}
\end{equation}
and recognizing the Carter separability constant $k$ we obtain the polar equation
\begin{equation}
k=\left(\frac{\partial S_{\theta}}{\partial
\theta}\right)^{2}+csc^{2}\theta(L+qQ\cos\theta)^{2}~.\label{i.7}
\end{equation}
 Finally, we find  formal solutions for the radial and polar
components of the action
\begin{equation}
S_{r}(r,k)= \epsilon \int \sqrt{\left(E-
{g\,Q\over r} \right)^{2}-
 \frac{f(r)}{a(r) }\left(m^2+\frac{k}{r^{2}}\right)}\,
\frac{dr}{f(r)} ~,\label{i.9}
\end{equation}
\begin{equation}
S_{\theta}(\theta,k,L)= \epsilon \int
\sqrt{k-csc^{2}\theta(L+qQ\cos\theta)^{2}}\,d\theta~,\label{i.9}
\end{equation}
where $\epsilon=\pm 1$.

Now,  considering
${\delta S\over\delta k}=0$,
${\delta S\over\delta m^2}=0$, ${\delta S\over\delta E}=0$ and ${\delta S\over\delta L}=0$, and from the Hamilton-Jacobi method, we simplify our
study to the following quadrature problem
\begin{equation}
\int
{d\theta\over \sqrt{k-csc^{2}\theta(L+qQ\cos\theta)^{2}}}=
\int {dr\over r^2\,a(r)\,\sqrt{\left(E-
{g\,Q\over r} \right)^{2}-
 \frac{f(r)}{a(r) }\left(m^2+\frac{k}{r^{2}}\right)}}~,
\end{equation}
\begin{equation}
\tau (r)=\epsilon \int{dr\over a(r)\,\sqrt{\left(E-
{g\,Q\over r} \right)^{2}-
 \frac{f(r)}{a(r) }\left(m^2+\frac{k}{r^{2}}\right)}}~,
\end{equation}

\begin{equation}\label{t}
t(r)=\epsilon \int
{\left[E-
{g\,Q\over r} \right]\,dr\over f(r)\,
\sqrt{\left(E-
{g\,Q\over r} \right)^{2}-
 \frac{f(r)}{a(r) }\left(m^2+\frac{k}{r^{2}}\right)}}~,
\end{equation}
\begin{equation}
\phi(r)=\epsilon \int
{ csc^{2}\theta(L+qQ\cos\theta)\,dr\over \,
\sqrt{k-csc^{2}\theta(L+qQ\cos\theta)^{2}}}~.\label{i.10}
\end{equation}
Defining the Mino time $\gamma$ as $r^2\,d\gamma=d\tau$,
we can express the equations of motion in terms of the new time parameter
\begin{equation}\label{theta}
\left({d\theta\over d\gamma}\right)^2=
k-csc^{2}\theta(L+qQ\cos\theta)^{2}~,
\end{equation}
\begin{equation}\label{t21}
\left({d\,r\over d\gamma}\right)^2=(r-Q^2/M)\left[
(r-Q^2/M)\left(E\,r-g\,Q \right)^{2}-
 (r-2M)\left(m^2\,r^{2}+k\right)\right]~,
\end{equation}
\begin{equation}\label{e16}
{d\phi\over d\gamma}=csc^{2}\theta(L+qQ\cos\theta)~,
\end{equation}
\begin{equation}
{d\,t\over d\gamma}=r^{2}\left({r-Q^2/M\over r-2M}\right)\left[E-{g\,Q\over r}\right]~.\label{time}
\end{equation}
In this way we have the equations of motion for our magnetically charged particle moving in the background of stringy magnetically black hole. In  the next section will perform a general analysis of the equations of motion.

\section{General analysis of the equations of motion}
\label{eqofmot}

Considering the Hamilton-Jacobi equations (\ref{theta})-(\ref{time}), the constants of motion (energy, angular momentum and separation constant), the parameters of the metric and the charges of the test particle we will analyze and solve the equations of motion that  characterize the various  types of orbits.

\subsection{ Analysis of the angular motion ($\theta$-motion) }

In order to study the $\theta$-motion, we consider the equation of motion (\ref{theta}), which can be rewritten using the Mino time as  $d \theta/d
\gamma=\sqrt{\Theta}$, where the coordinate $\theta$ is a polar angle that can take only positive values. Then,
\begin{equation}
\Theta=
k-csc^{2}\theta(L+qQ\cos\theta)^{2} \geq0~,
\end{equation}
where the separability constant is definite positive. Now, thought the change of variables $\xi=\cos\theta$, equation (\ref{theta}) yields
\begin{equation}
\frac{d \xi}{d \gamma}=\sqrt{\Theta_{\xi}};\qquad
\text{with} \qquad
\Theta_{\xi}=k-L^{2}-2LqQ\xi-(k+q^{2}Q^{2})\xi^{2}~.
\label{h.1}
\end{equation}
The roots of the function $\Theta_{\xi}$ are given by
\begin{equation}
\theta_1=\cos^{-1}\left[{LqQ\over k+q^{2}Q^{2}}
\left(\sqrt{1+{(k-L^2)(k+q^{2}Q^{2})\over L^2q^{2}Q^{2}}}-1\right)\right]~,
\end{equation}
\begin{equation}
\theta_2=\cos^{-1}\left[{LqQ\over k+q^{2}Q^{2}}
\left(-\sqrt{1+{(k-L^2)(k+q^{2}Q^{2})\over L^2q^{2}Q^{2}}}-1\right)\right]~,
\end{equation}
which define the cone's angles that confine the movement of the particle.  Then, $\gamma(\theta)$ yields
\begin{equation}\gamma(\theta)=\frac{1}{\sqrt{k+q^{2}Q^{2}}}
\arccos\left[\frac{(k+q^{2}Q^{2})\cos\theta+LqQ}{\sqrt{(k-L^2)(k+q^{2}Q^{2})+L^2q^{2}Q^{2}}}\right]
,\label{h.2}\end{equation}
where we have used that $\gamma_{0}=0$ for $\theta_{0}=\theta_{1}$.
Also, the above equation can be inverted, which yields
\begin{equation}\theta(\gamma)=
\arccos\left[\frac{\sqrt{(k-L^2)(k+q^{2}Q^{2})+L^2q^{2}Q^{2}}
\cos(\sqrt{k+q^{2}Q^{2}}\,\gamma)-LqQ}{k+q^{2}Q^{2}}\right]~,
\label{h.3}\end{equation}
for $k>L^2$.

\subsection{Analysis of the radial motion ($r$-motion) }
\label{radial}

Now, we consider the motion of the particle with respect to the $r$-coordinate. We will focus on equation (\ref{t}) in order to obtain the velocity of the particle $dr/dt$. The condition of  turning point $(\frac{dr}{dt})_{r=r_t}=0$ allows us to define
\begin{equation}
\left(E-{g\,Q\over r} \right)^{2}-
{f(r)\over a(r)}\left(m^2+\frac{k}{r^{2}}\right)=(E-V_-)(E-V_+)~,\label{i.12}
\end{equation}
where we can recognize the effective potential for the particle
with mass $m$ and magnetic  charge $g$ as
\begin{equation}
V_{\pm}(r)={g\,Q\over r}\pm\sqrt{{f(r)\over a(r)}\left(m^2+\frac{k}{r^{2}}\right)}~.\label{i.13}
\end{equation}
Since the negative branches have no classical interpretation,  they are associated with antiparticles in the
framework of quantum field theory \cite{Deruelle:1974zy}. We choose the positive branch of the effective potential
$V_{eff}=V_{+}\equiv V$.

\begin{figure}[!h]
\begin{center}
\includegraphics[width=80mm]{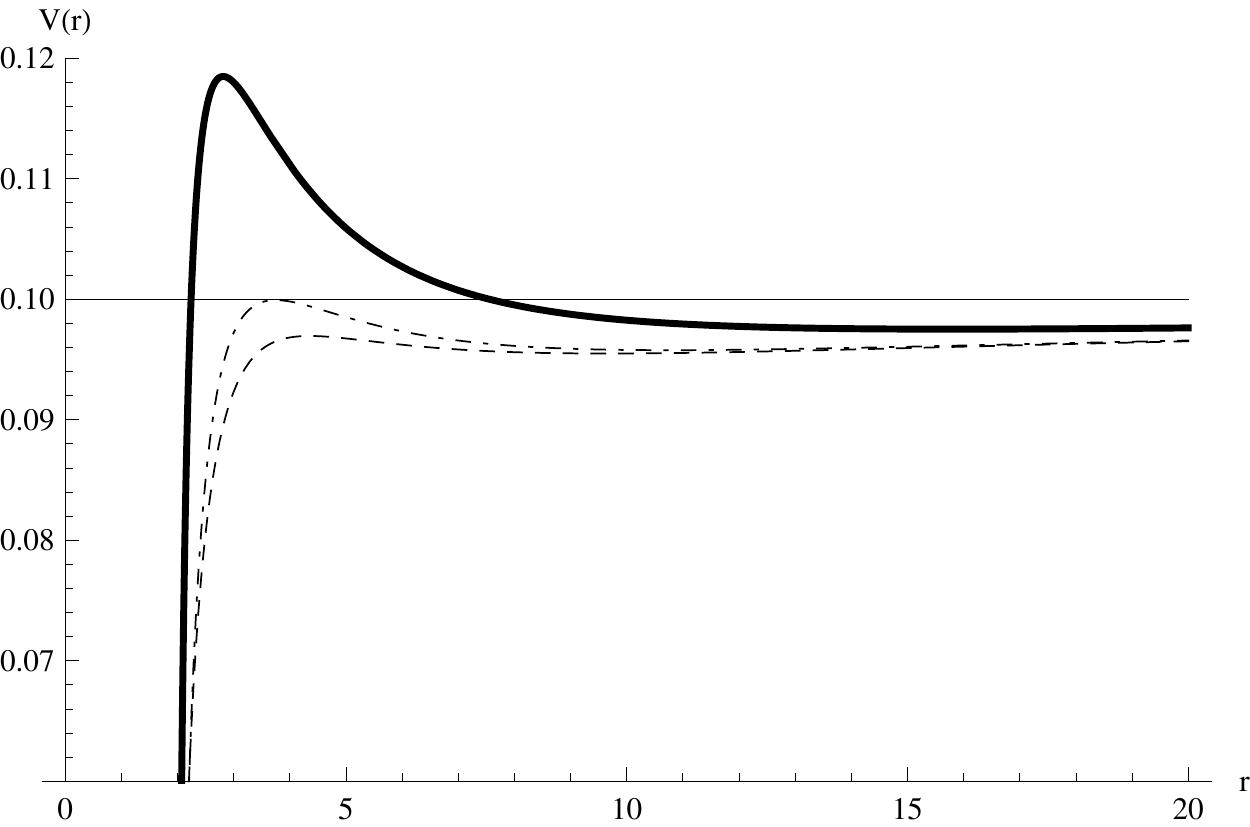}
\end{center}
\caption{Effective potential
 for $g=-0.05$, $M=1$, $m=0.1$, and different values of the magnetic charge $Q=0$ (dashed line), $Q=Q_c\approx 1.02$ (dot-dashed line), $Q=1.3$ (thick line) and $E=m$ (thin horizontal line).}
\label{Qcritic}
\end{figure}

Now, for simplicity we will consider the case where $Qg \leq 0$. Note that  $V_{eff}(r\rightarrow \infty)=m$.
So, we plot in Fig. \ref{Qcritic}
 the effective potential for different values of the magnetic charge $Q$ with $g=-0.05$, $M=1$, $m=0.1$.
 We define the the  bound and unbound orbits if along the orbits $r$ remains bounded or not.
  We can observe that there is a critical magnetic charge $Q_c$, where the energy of the unstable circular orbit takes the value $m$. For a $Q<Q_c$, the effective potential show that all the unbound trajectories ($E\geq m$) can fall to the horizon or can escape to the infinity. If $Q>Q_c$, the maximum value of the effective potential is greater than $m$ and the particles with energy $m<E<E_c$ have return points, the magnetic particles located in the right side of the potential that arrive from the infinity have a point of minimum approximation and it is scattered to the infinity which we can call "Magnetic Rutherford scattering". However, for particles located in the left side of the potential have a return point from which plunges to the horizon.

\begin{figure}[!h]
\begin{center}
\includegraphics[width=80mm]{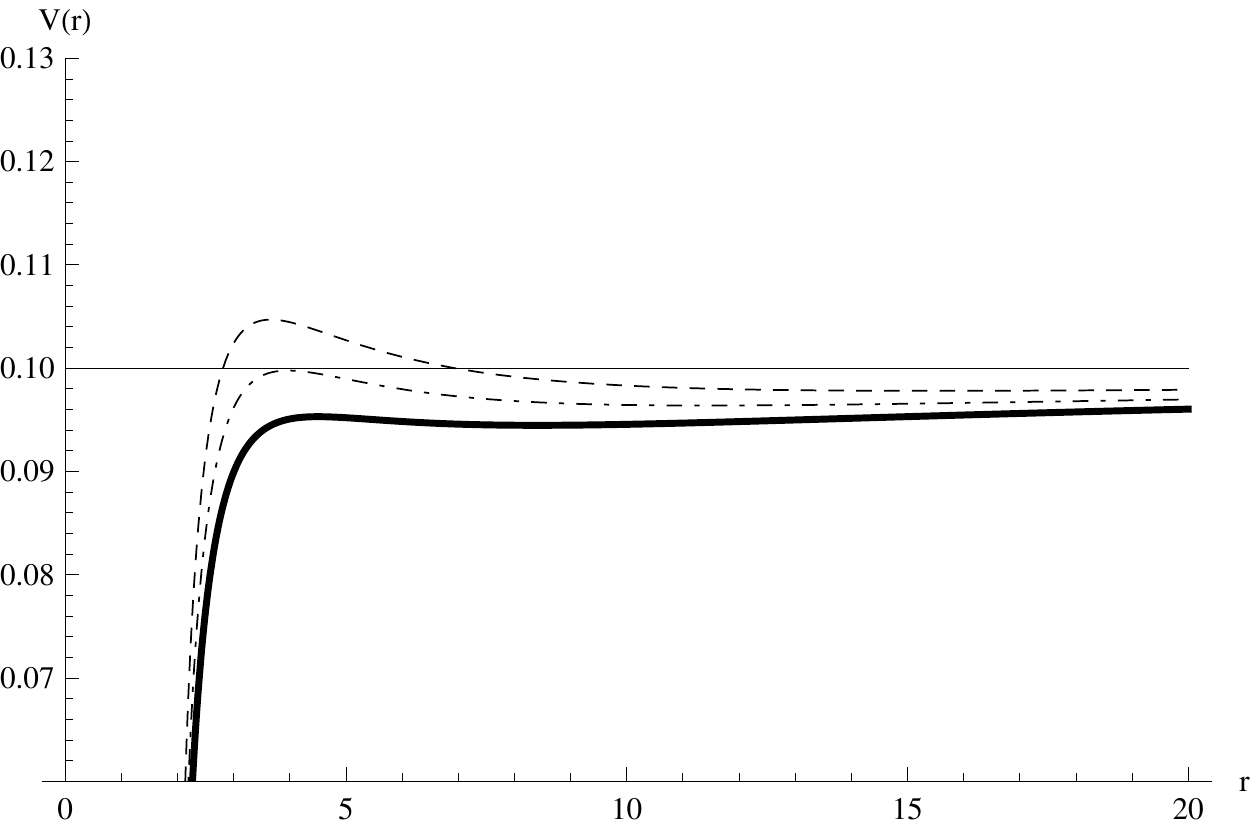}
\end{center}
\caption{Effective potential
 for $M=1$, $m=0.1$, $Q=0.75$ and different values of the magnetic charge of the test particle $g=0$ (dashed line), $g=g_c\approx -0.025$ (dot-dashed line), $g=-0.05$ (thick line) and $E=m$ (thin horizontal line).}
\label{gcritic}
\end{figure}

  Then, in Fig. \ref{gcritic} we plot the effective potential for different values of the magnetic charge of the test particle $g$ with $M=1$, $m=0.1$, $Q=0.75$. Here, we can observe that there a critical value of the magnetic charge of the test particle $g_c$ where the energy of the unstable circular orbit takes the value $m$. For a $g>g_c$, the effective potential shows that the test particle appears in the scattering zone, while  for $g<g_c$ does not appear in the scattering zone.

\begin{figure}[!h]
\begin{center}
\includegraphics[width=80mm]{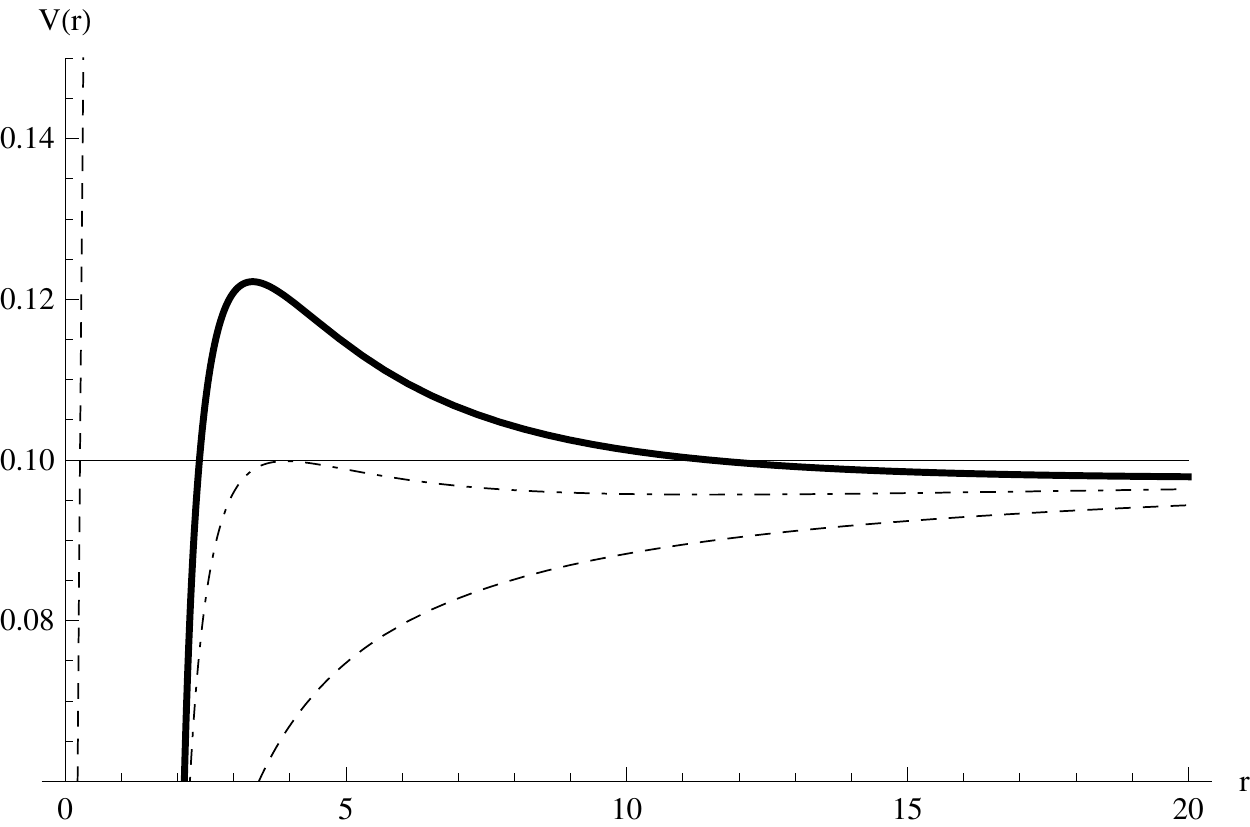}
\end{center}
\caption{Effective potential
 for $g=-0.05$, $M=1$, $m=0.1$ and different values of the Carter separability constant $k=0$ (dashed line), $k=k_c\approx 0.168$ (dot-dashed line), $k=0.3$ (thick line) and $E=m$ (thin horizontal line).}
\label{kcritic}
\end{figure}

 Finally in Fig. \ref{kcritic} we plot the effective potential for different values of the Carter separability constant $k$ with $g=-0.05$, $M=1$, $m=0.1$. We can observe that the behaviour is similar to the first case, that is, there a critical value of the Carter separability constant $k_c$, where the energy of the unstable circular orbit takes the value $m$. For a $k<k_c$, the effective potential shows that all the unbound trajectories ($E\geq m$) can fall to the horizon or can escape to the infinity. If $k>k_c$, the maximum value of the effective potential is greater than $m$ and the particles with energy $m<E<E_c$ have a return points, the magnetic particles located in the right side of the potential that arrive from the infinity have a point of minimum approximation and it is scattered to the infinity which we have called "Magnetic Rutherford scattering". However, for particles located in the left side of the potential have a return point from which plunge to the horizon.


\newpage
Let us  rewrite the radial equation (\ref{t21}) as
\begin{equation}\label{tl11a}
  \left(\frac{dr}{d\gamma}\right)^2=\left(r-{Q^2\over M} \right)\,R(r)~,
\end{equation}
where we define
$$
R(r)\equiv
(r-Q^2/M)\left(E\,r-g\,Q \right)^{2}-
 (r-2M)\left(m^2\,r^{2}+k\right)~,
$$
now this function can be written as the characteristic polynomial
\begin{equation}
R(r)=a_3 r^3 + a_2 r^2 + a_1 r+a_0~,
\end{equation}
where
\begin{eqnarray}\label{c12}
a_0 &=&2kM-{g^2Q^4 \over M}~,\qquad\qquad\qquad\,\,\,\,\,
a_1 =-k+g^2Q^2+{2gQ^3E^2\over M}~,\\
a_2 &=&2m^2M-2gQE-{Q^2E^2\over M}~,\qquad
a_3 =E^2-m^2~.\qquad
\end{eqnarray}
Now, by performing a  change of variables
\begin{equation}\label{cv}
U(r)={1\over 4(r-\alpha)}+{\beta_1\over 12}~,
\end{equation}
we obtain
\begin{equation}\label{c11}
r(\gamma)=\left({Q^2\over M}\right)+{1\over 4\,\wp(\omega_0\mp\sqrt{b_0}\,\gamma; g_{2},g_{3})-
\beta_1/3}~,
\end{equation}
where $\wp(x; g_2, g_3)$ is the $\wp$-Weierstrass elliptic function, with
the Weierstrass invariant given by
\begin{equation}\label{c12}
g_2 ={\beta_{1}^{2} \over12}-{\beta_{2} \over 4},\qquad
g_3 ={\beta_{1}\beta_{2} \over 48}-{\beta_{1}^{3} \over 216}
-\beta_{3}~,\qquad
\textrm{and}\quad
\omega_0=\wp^{-1}\left[{ 1\over 4(r_0-Q^2/M)}+{ \beta_{1}\over 12}\right]~,
\end{equation}
where
\begin{equation}
\beta_1 ={b_1\over b_0}~,\qquad
\beta_2 ={b_2\over b_0}~,\qquad
\beta_3 ={a_3\over b_0}~,
\end{equation}
and $r_0$ corresponds to an initial  arbitrary distance and
\begin{eqnarray}
\nonumber b_0 &=&a_0+a_1\left({Q^2\over M}\right)+a_2\left({Q^2\over M}\right)^2+a_3\left({Q^2\over M}\right)^3~,\\
\nonumber b_1&=&a_1+2a_2\left({Q^2\over M}\right)+3a_3\left({Q^2\over M}\right)^2~,\\
b_2&=& a_2+3a_3\left({Q^2\over M}\right)~.
\end{eqnarray}

\subsection{Analysis of the angular motion ($\phi$-motion) }

In order to obtain the $\phi$-motion, we consider Eq. (\ref{theta}) and Eq. (\ref{e16}) which allow us to write
\begin{equation} \phi(\theta)=\int^{\theta}_{\theta_{1}}
\frac{(L+qQ\cos\theta)d\theta}{\sin^{2}\theta\sqrt{\Theta}}~,\label{h.4}\end{equation}
which yields
\begin{eqnarray}
\nonumber \phi(\theta)&=&\frac{1}{2}\arccos
\left(\frac{(k-LqQ+q^{2}Q^{2})(1+\cos\theta)-(L-qQ)^2}
{(1+\cos\theta)\sqrt{k(k-L^2+q^{2}Q^{2})}}\right)+\\
&&\frac{1}{2}\left(
\arcsin
\left(\frac{(k+LqQ+q^{2}Q^{2})(1-\cos\theta)-(L+qQ)^2}
{(1-\cos\theta)\sqrt{k(k-L^2+q^{2}Q^{2})}}\right)+\frac{\pi}{2}\right)
~,\label{h.5}\end{eqnarray}
where $L>qQ$,  and we have used as initial condition
$\phi(\theta_{1})=0$ for simplicity. Then, it is possible to obtain $\phi(\gamma)$ by replacing $\theta(\gamma)$ in the above expression.

\subsection{ Analysis of the time motion ($t$-motion) }
Now, in order to describe the time motion we rewrite Eq. (\ref{t}) as
\begin{equation}
t(r)=\int_{r_0}^{r}r^2\left({r-\alpha\over r-2M}\right)
\left(E-{gQ\over r}\right){d\,r\over \sqrt{(r-\alpha)R(r)}}~,
\end{equation}
and using the same change of variables that we have used in the radial motion (\ref{cv}), we obtain the following solution
\begin{equation}
t(r)=A_1\left[F_1(r)-F_1(r_0)\right]+A_2\left[F_2(r)-F_2(r_0)\right]+
A\left[F(r)\right]~,
\end{equation}
where

$$
A_1={-gQ+E(2\alpha+\beta)\over 4\sqrt{b_0}}~,
\quad
A_2={gQ(\alpha+\beta)-E(\alpha+\beta)^2\over 4\beta\sqrt{b_0}}~,
\quad
A=E~,
$$

\begin{equation}
F_i(r)=\frac{1}{\wp^{'}(\Omega_i)}
\left[\zeta(\Omega_i)\wp^{-1}(U)
+\ln\left|\frac{\sigma[\wp^{-1}(U)-\Omega_i]}
{\sigma[\wp^{-1}(U)+\Omega_i]}\right|
\right]~,\end{equation}
with

$$
\Omega_1=\wp^{-1}\left({\beta_1\over 12}\right)~,
\quad
\Omega_2=\wp^{-1}\left({3+\beta\beta_1\over 12\beta}\right)~,
$$
and
\begin{equation}
F(r)=A_0\left(\varPi-\varPi_0\right)-
\sum^{2}_{i=1}
\frac{1}{(\wp^{'}[\omega_i])^2}
\left[\zeta[\varPi-\omega_i]-\zeta[\varPi_0-\omega_i]
+\frac{\wp^{''}[\omega_i]}{\wp^{'}[\omega_i]}
\ln\left|\frac{\sigma[\varPi-\omega_i]}
{\sigma[\varPi_0-\omega_i]}\right| \right]~,
\end{equation}
with
$$
\varPi\equiv\wp^{-1}(U) \quad \text{and} \quad
{\beta_1\over 12}={1\over (\wp[\omega_1]-{\beta_1\over 12})^2}
={1\over (\wp[\omega_2]-{\beta_1\over 12})^2}~,
$$
and
$$
A_0=-
\sum^{2}_{i=1}
\left[\frac{\wp[\omega_i]}{(\wp^{'}[\omega_i])^2}
+\frac{\wp^{''}[\omega_i]\zeta[\omega_i]}{(\wp^{'}[\omega_i])^3} \right]~,
$$
where we have used $\varPi_0=\wp^{-1}[U(r_0)]$.

\section{The orbits}
\label{orbits}

\begin{figure}[!h]
\begin{center}
\includegraphics[width=100mm]{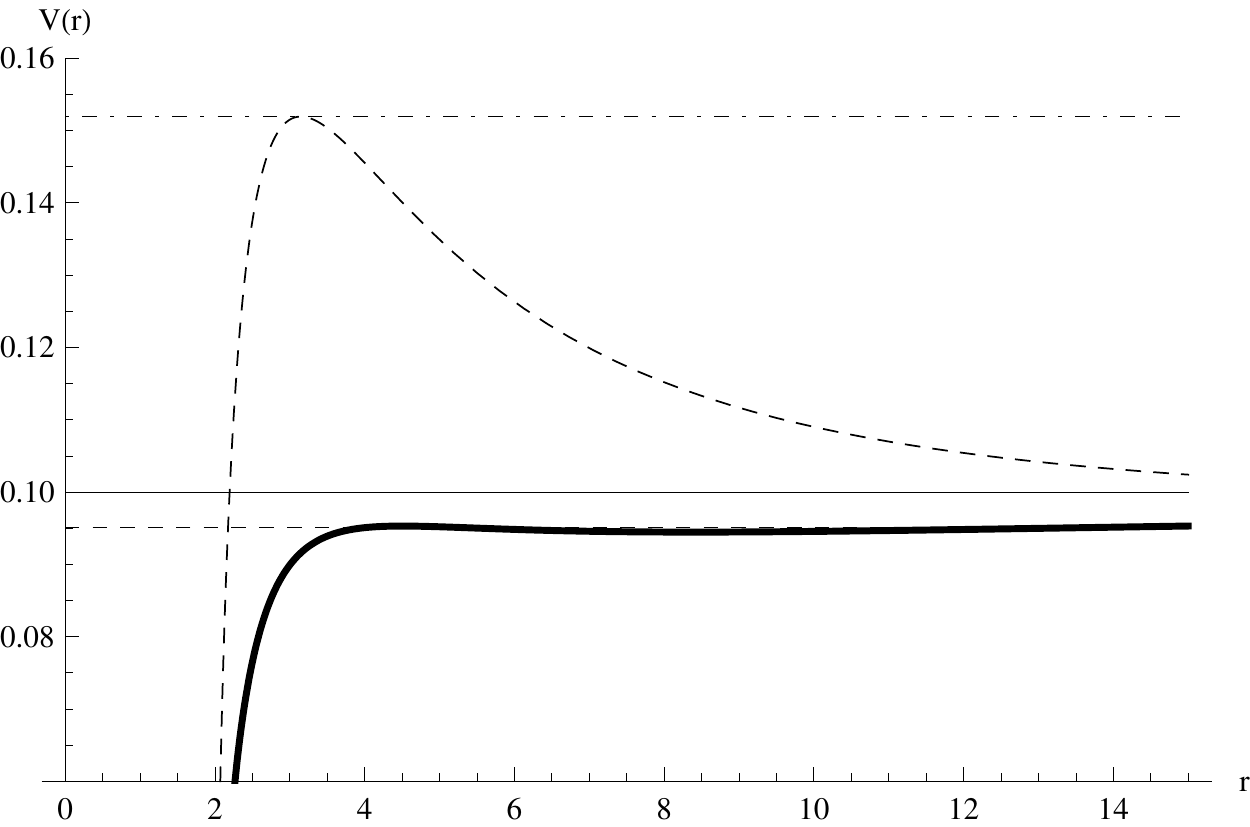}
\end{center}
\caption{Effective potential for bound orbits $k=0,14$ (thick line), unbound orbit $k=0,5$ (dashed line), $E=0.0951$, $E\approx 0.1519$, and $E=m$ with $g=-0.05$, $Q=0.75$, $M=1$, $m=0.1$.}
\label{f1}
\end{figure}

In this section we analyze the different kind of orbits that the test particles follow  in the magnetically charged stringy black hole for the following parameters $m=0.1$, $q=0.05$, $g=-0.05$, $M=1$ and $Q=0.75$. The potential is shown in Fig. \ref{f1},
where we observe the two possible cases, one of them is the potential for bound orbits ($E<m$) and the other one is the potential for unbound orbits
($E\geq m$). The orbits of the first kind are the relativistic analogues of the Keplerian orbits to which they tend in the Newtonian limit. The orbits of the second kind have no Newtonian analogues \cite{chandra}.

\begin{figure}[!h]
\begin{center}
\includegraphics[width=60mm]{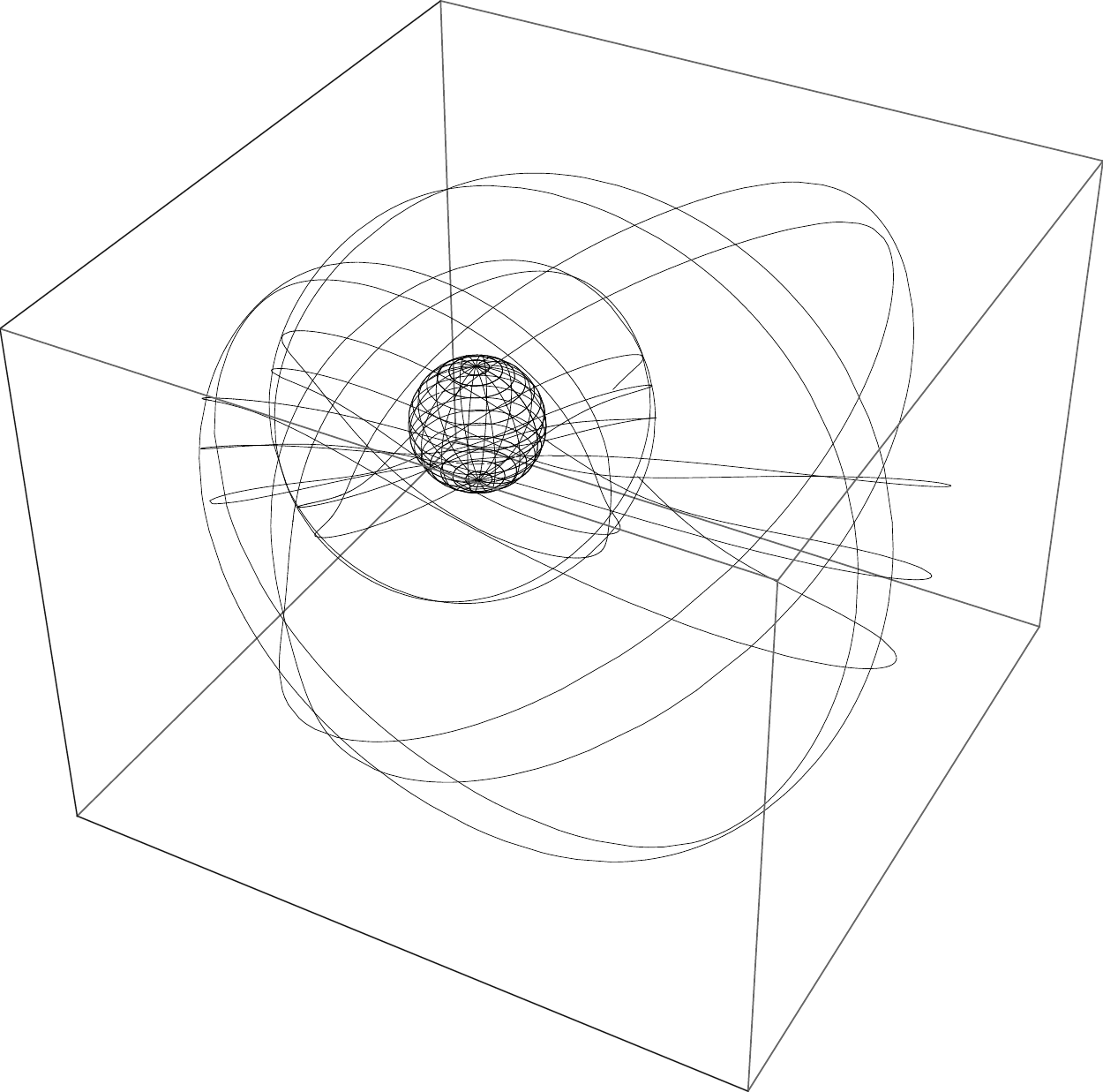}\quad\quad\quad\quad
\includegraphics[width=60mm]{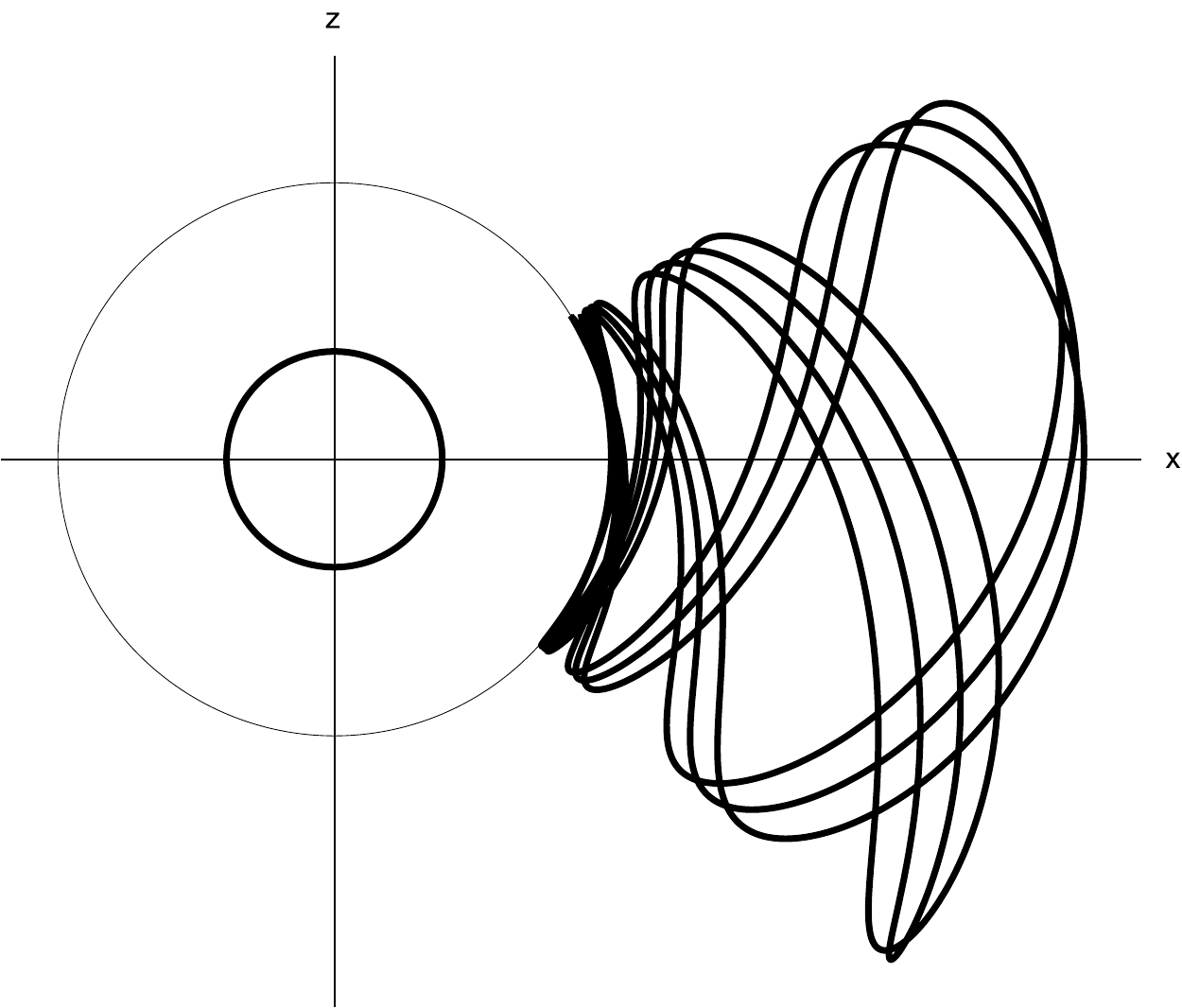}
\end{center}
\caption{Planetary orbit with $k=0.14$, $L=0.3$ and $E=0.0951$. Left figure for three-dimensional motion, where the central circle correspond to the event horizon, and right figure is the projection of the orbit onto the $x-z$-plane, where the small circle corresponds to the event horizon and the other one corresponds to the periastro distance.}
\label{f5}
\end{figure}

\begin{figure}[!h]
\begin{center}
\includegraphics[width=60mm]{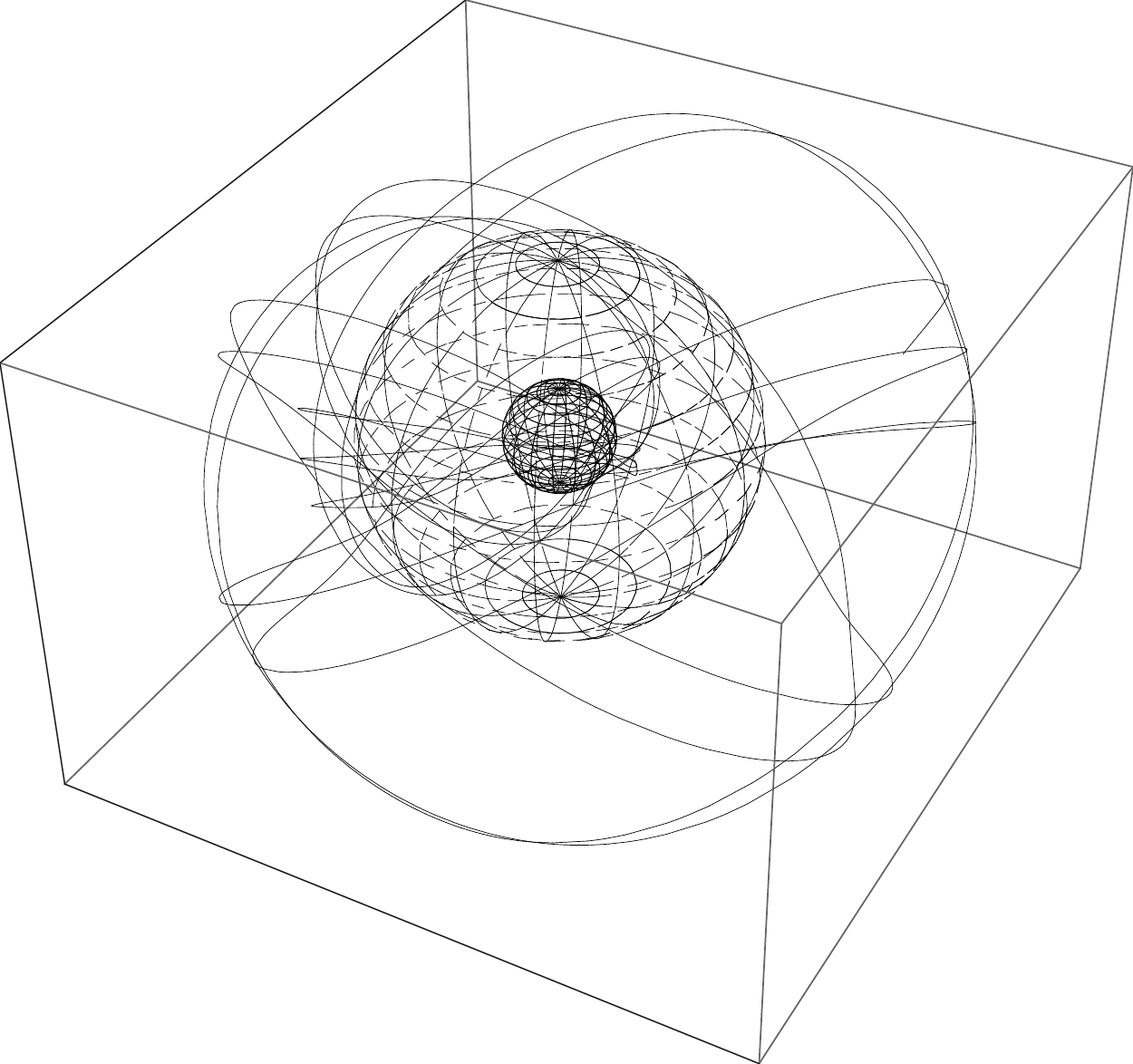}\quad\quad\quad\quad
\includegraphics[width=60mm]{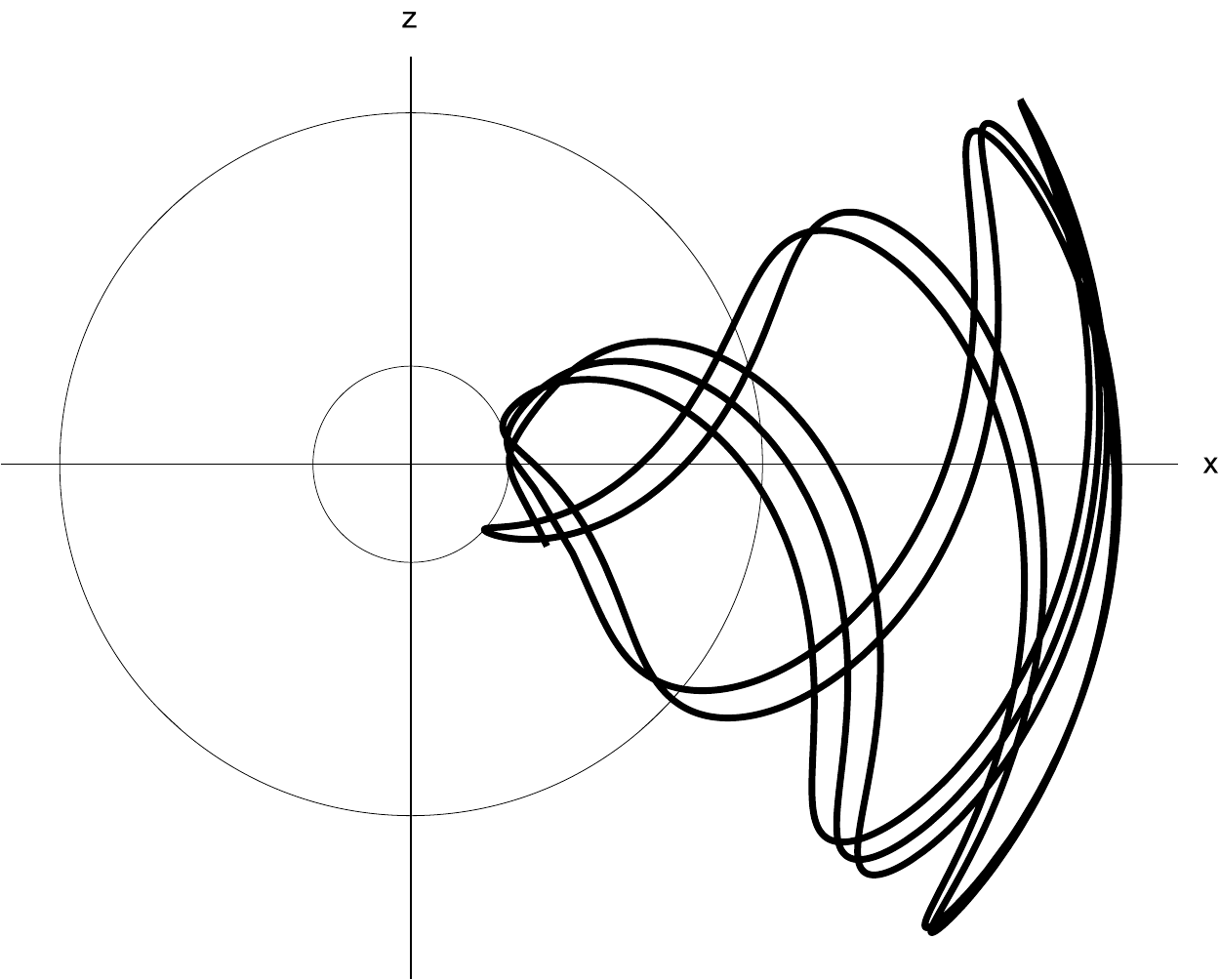}
\end{center}
\caption{Bound orbits of second kind with $k=0.14$, $L=0.3$ and $E=0.0951$. Left figure for three-dimensional motion, where the central circle corresponds to a return point inside the horizon (by simplicity we choose $Q^2/M$, as return point), and right figure is the projection of the orbit onto the $x-z$-plane, where the small circle corresponds to the return point  and the other one corresponds to event horizon.  Note that the trajectory has a physical meaning outside the horizon.}
\label{f6}
\end{figure}

\begin{figure}[!h]
\begin{center}
\includegraphics[width=60mm]{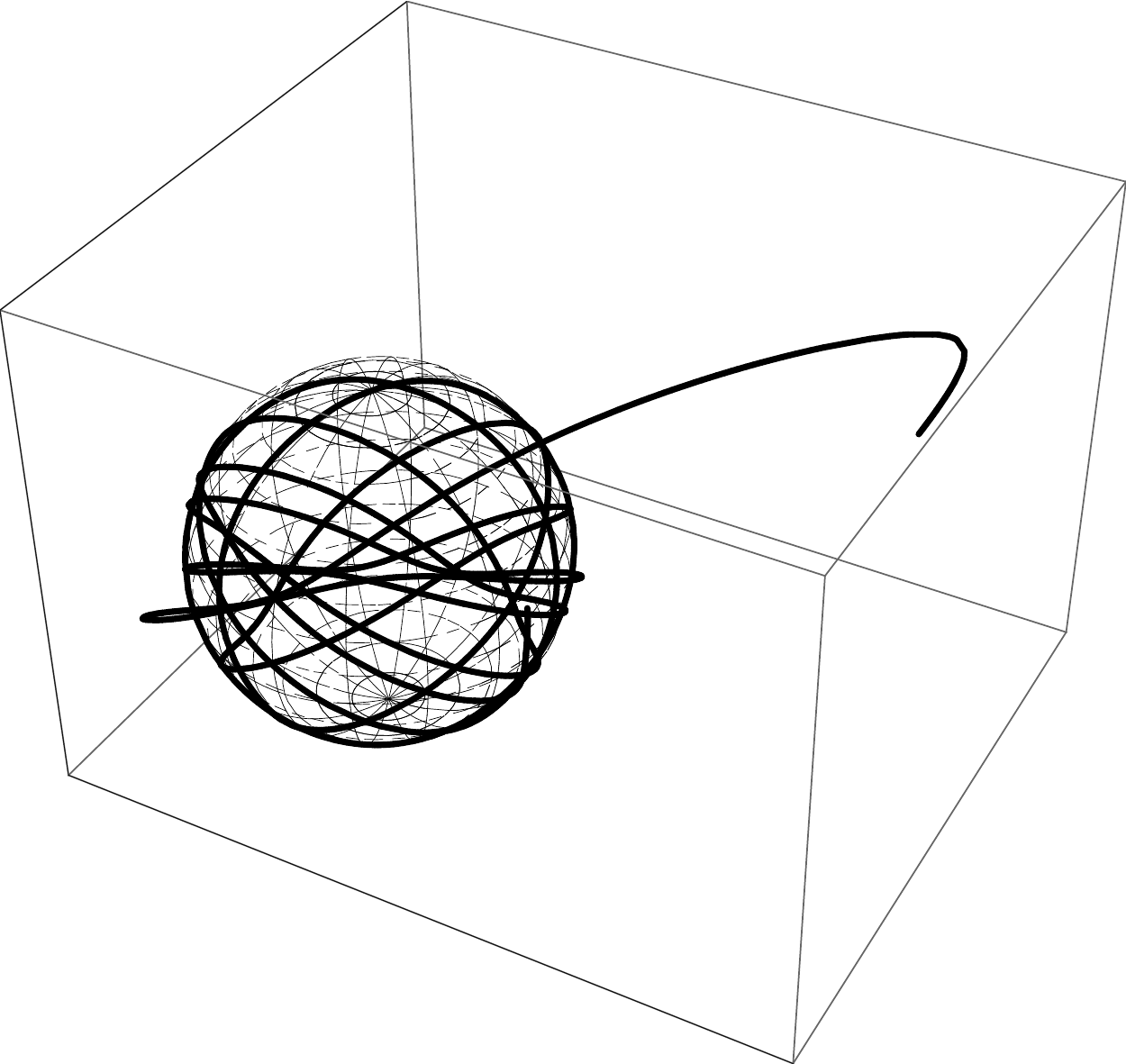}\quad\quad\quad\quad
\includegraphics[width=60mm]{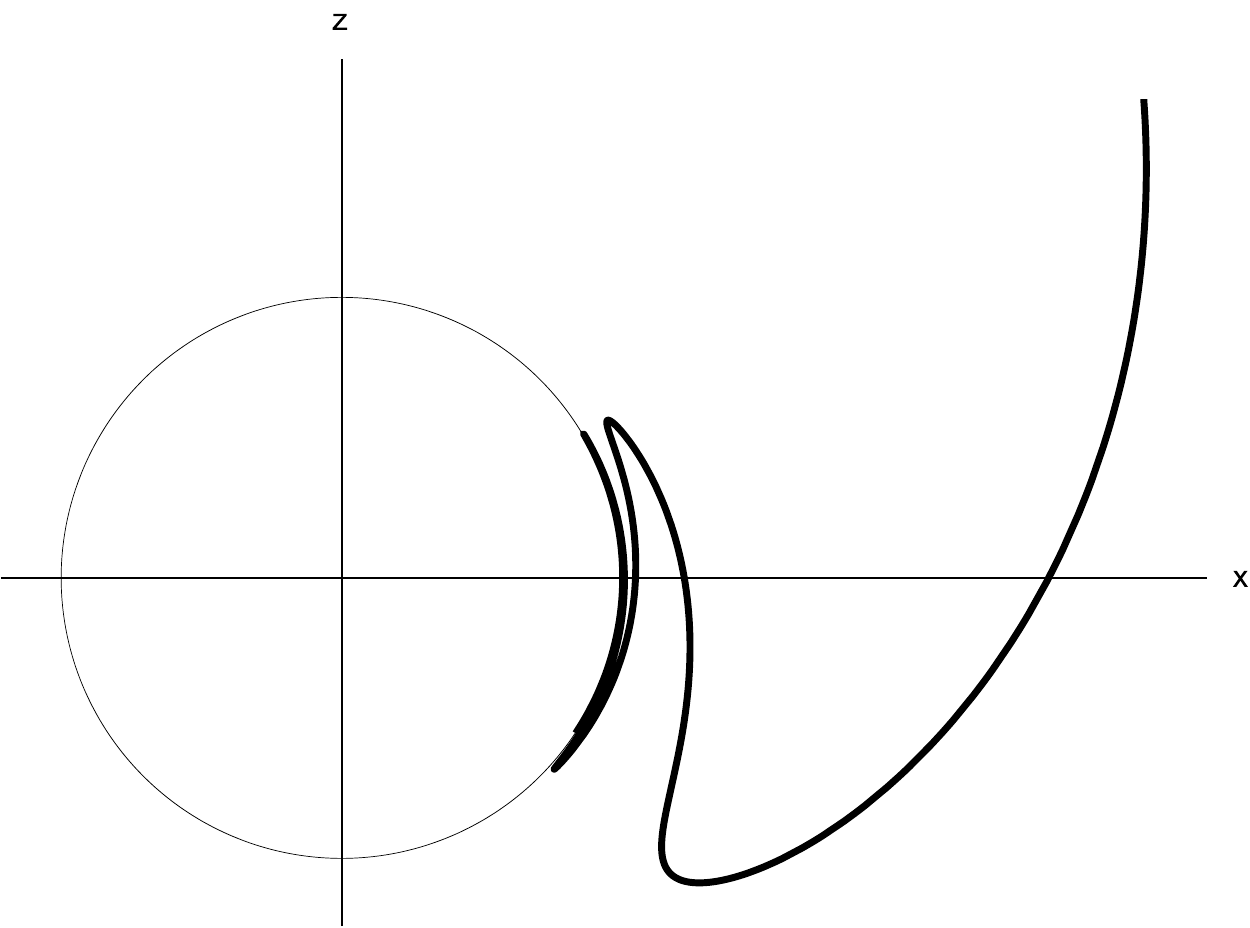}
\end{center}
\caption{Critical orbits of first kind with $k=0.14$, $L=0.3$ and $E=0.0953$. Left figure for three-dimensional motion, and right figure is the projection of the orbit onto the $x-z$-plane. The circle corresponds to the unstable circular orbit.}
\label{f7}
\end{figure}

In order to obtain a full description of the radial motion of charged particles,
we study separately the two possible cases.
For bound orbits, in Fig. \ref{f5} we show the behaviour of the orbit for a charged particle, that oscillate between the periastro and apoastro distance, this first kind of orbit corresponds to a planetary orbit. Also, we show the projection of the orbit onto the $x-z$-plane. Then, in Fig. \ref{f6} we show the orbit of a particle with the same energy that the previous case, but it is localized to left of the potential. Also, we show the projection of the orbit onto the $x-z$-plane it is an orbit of the first kind. Then, in Fig. \ref{f7} we show the behaviour of critical orbits of first kind with $E=0.0953$, we can observe that the trajectory starts at a certain aphelion distance and approaches the circle, asymptotically, by spiralling around it an infinite number of times. Also, we show the projection of the orbit onto the $x-z$-plane.

\begin{figure}[!h]
\begin{center}
\includegraphics[width=60mm]{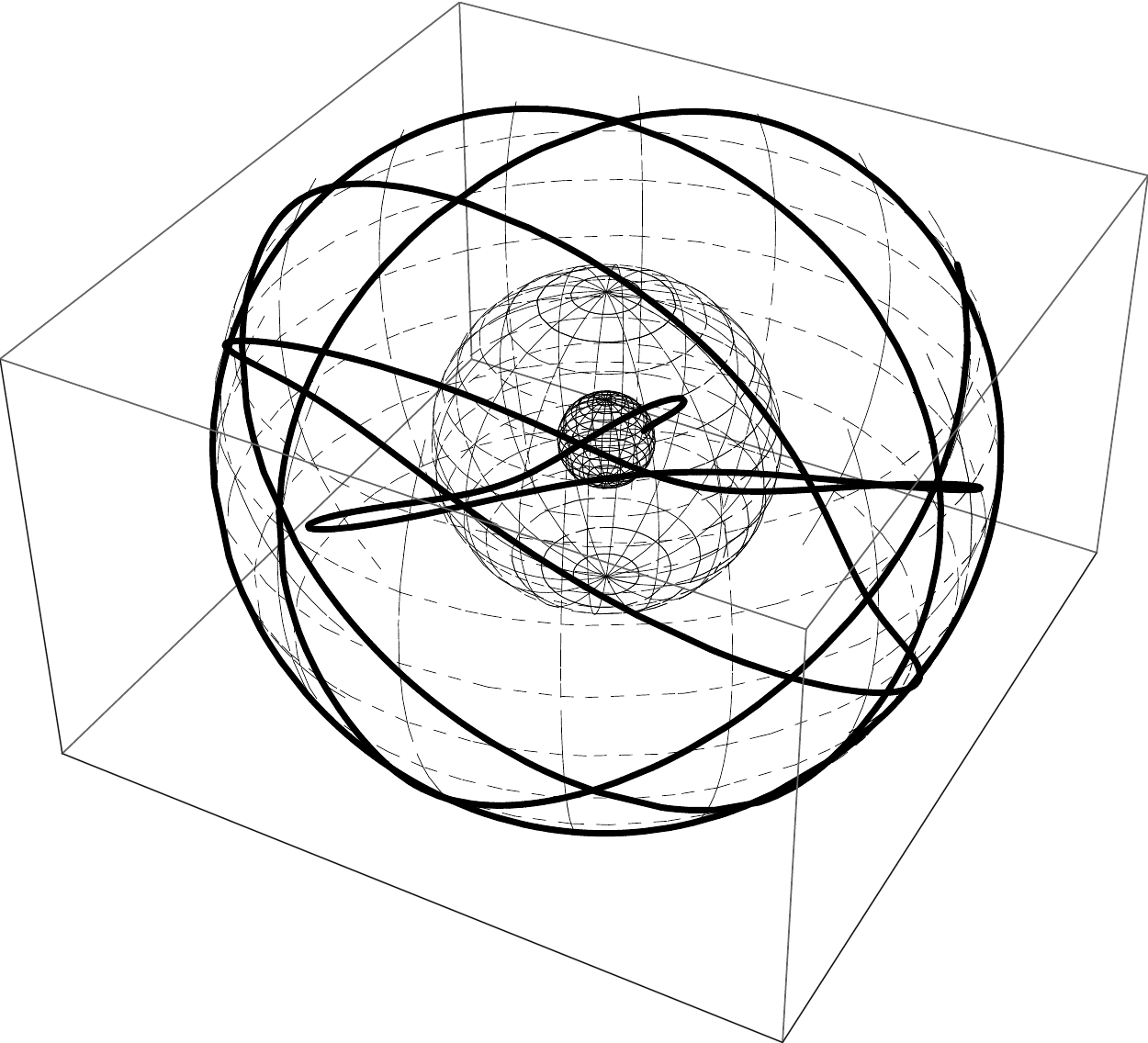}\quad\quad\quad\quad
\includegraphics[width=60mm]{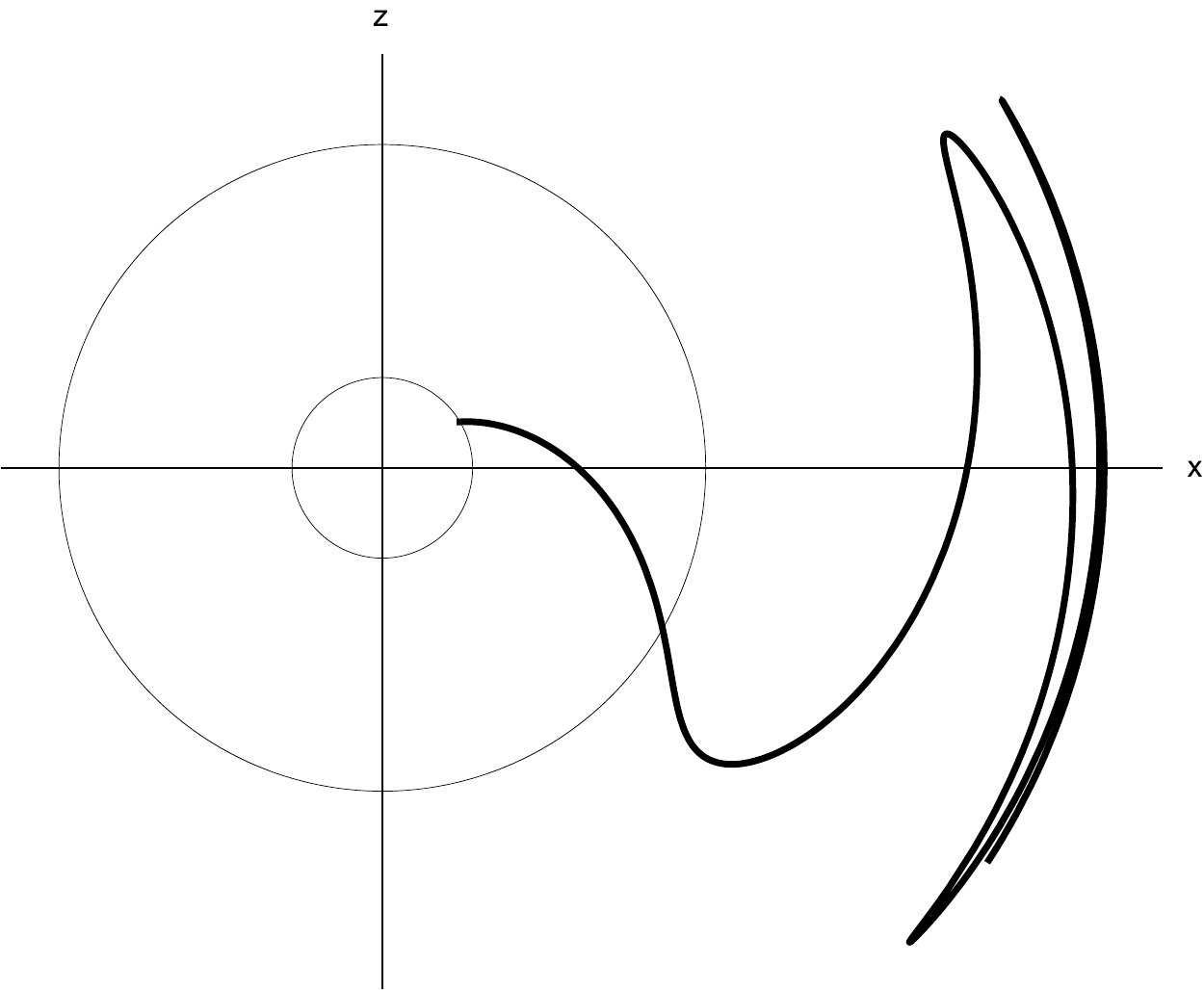}
\end{center}
\caption{Critical orbits of second kind with $k=0.14$, $L=0.3$ and $E=0.0953$. Left figure for three-dimensional motion, where the central circle corresponds to a return point inside the horizon (by simplicity we choose $Q^2/M$, as return point), and right figure is the projection of the orbit onto the $x-z$-plane, where the small circle corresponds to the return point  and the other one corresponds to event horizon. Note that the trajectory has a physical meaning outside the horizon.}
\label{f8}
\end{figure}

\begin{figure}[!h]
\begin{center}
\includegraphics[width=60mm]{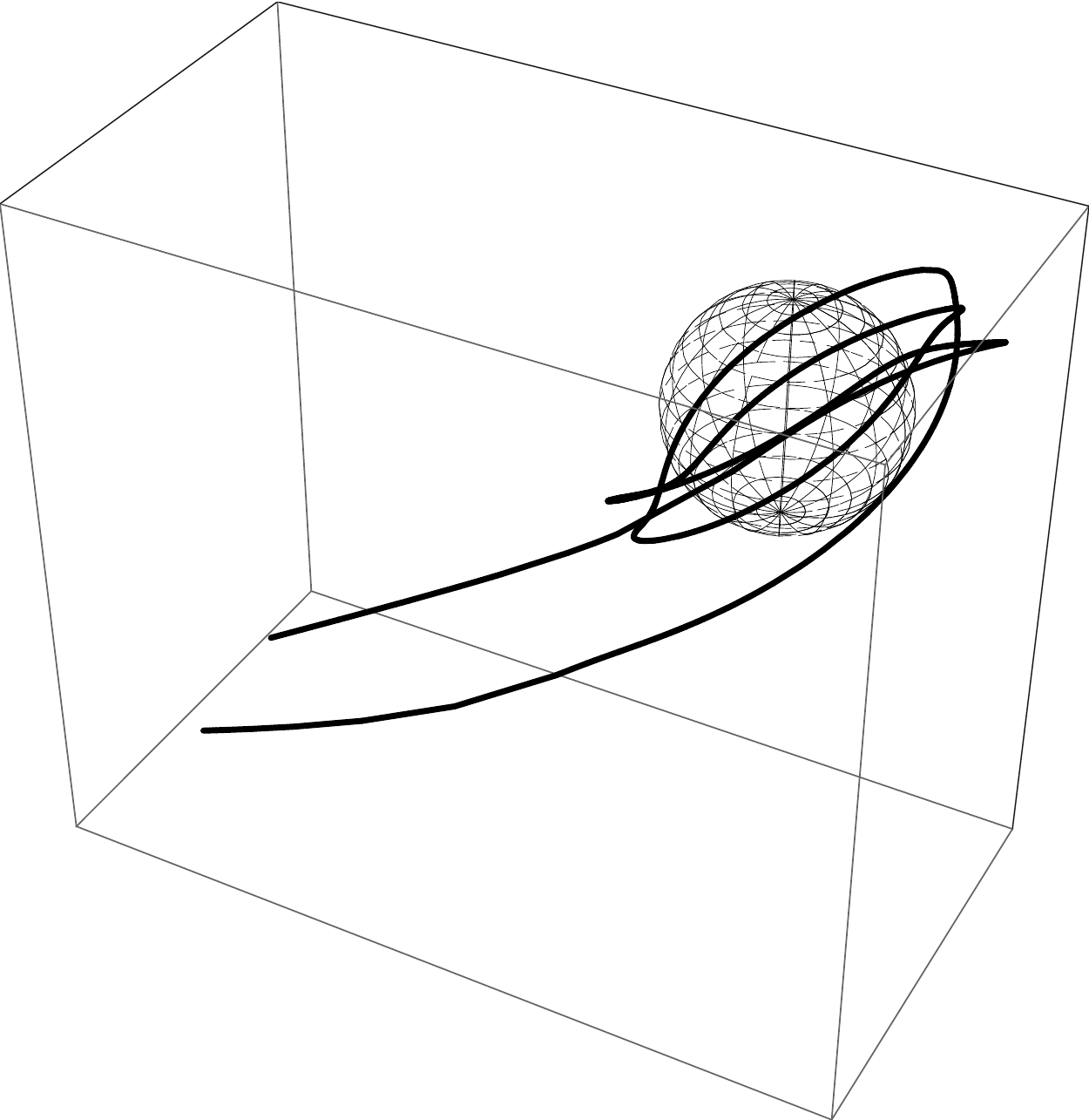}\quad\quad\quad\quad
\includegraphics[width=60mm]{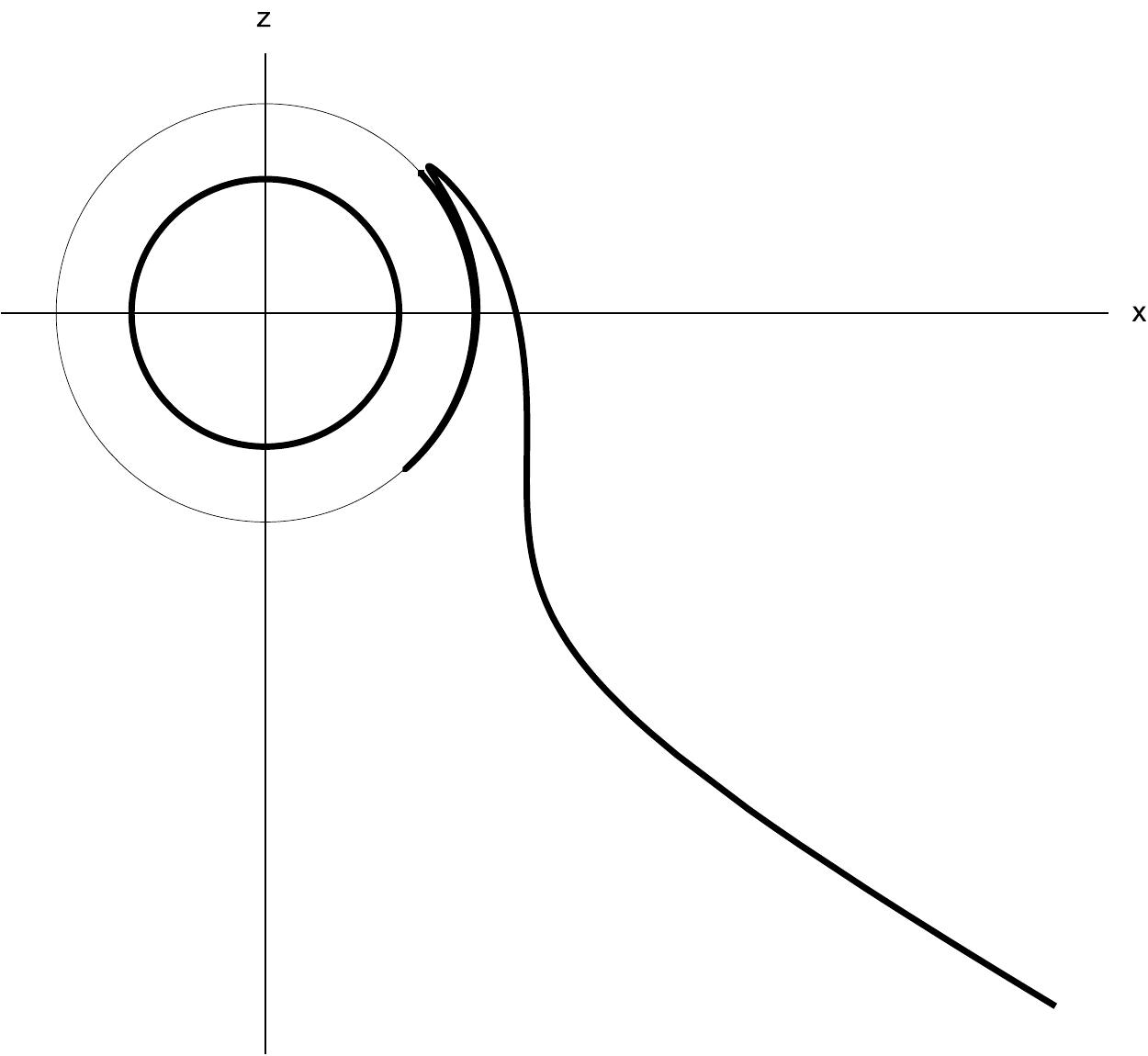}
\end{center}
\caption{Trajectory for a particle that arrive from infinity (Magnetic Rutherford scattering) with $k=0.5$, $L=0.3$ and $E=0.1519$. Left figure for three-dimensional motion, and right figure is the projection of the orbit onto the $x-z$-plane. The small circle correspond to the event horizon and the other one corresponds to  the closest  approach distance.}
\label{f9}
\end{figure}

In Fig. \ref{f8} we show the behaviour of critical orbits of second kind with the same energy that the previous case, but the particle is localized to left of the potential. Also, we show the projection of the orbit onto the $x-z$-plane.
On the other hand,  for unbound orbits in Fig. \ref{f9} we show the trajectory of a particle that arrive from infinity with $E=0.1519$, and we show the projection of the orbit onto the $x-z$-plane. It is an orbit of the first kind, due to the fact that the orbit is analogue to the hyperbolic orbit of the Newtonian theory \cite{chandra}.

\newpage

\section{The observables}
\label{observables}

In this section following  \cite{Kagramanova:2010bk} we calculate some possible observables. However, as our work is about massive particles we will study the perihelion shift and the  Lense-Thirring effect. So, we consider here bound orbits, like the planetary orbits,
these orbits precess between the aphelion distance, $r_A$, and the perihelion
distance, $r_P$. The $r$-motion is periodic with a
period given by
\begin{equation}
\omega_r=2\gamma(r_A)={2\over \kappa_p}
\left(\wp^{-1}\left[{ 1\over 4(r_P-Q^2/M)}+{ \beta_{1}\over 12}\right]
-\wp^{-1}\left[{ 1\over 4(r_A-Q^2/M)}+{ \beta_{1}\over 12}\right]\right)~.
\end{equation}The corresponding orbital frequency is $2\pi/\omega_r$. On the other hand, the period of the $\theta$-motion is given by
\begin{equation}
\omega_{\theta}=2\gamma(\theta_2)={2\pi\over \sqrt{k+q^2Q^2}}~,
\end{equation}
and the corresponding frequency by $2\pi/\omega_{\theta}$.

The secular accumulation rates of the angle $\phi$ and the time $t$ are given by
\begin{equation}
Y_{\phi}={2\over\omega_{\theta}}\phi(\theta_2)= \sqrt{k+q^2Q^2}~,
\end{equation}
\begin{equation}
\Gamma={2\over\omega_{r}}t(r_A)={2\over\omega_{r}}
\left(A_1\left[F_1(r_A)-F_1(r_P)\right]+A_2\left[F_2(r_A)-F_2(r_P)\right]+
A\left[F(r_A)\right]\right)~,
\end{equation}
and the orbital frequencies $\Omega_r$,  $\Omega_{\theta}$ and $\Omega_{\phi}$ read\begin{equation}
\Omega_r={2\pi\over\omega_{r}}{1\over\Gamma}~,\quad
\Omega_{\theta}={2\pi\over\omega_{\theta}}{1\over\Gamma}~,\quad
\Omega_{\phi}={Y_{\phi}\over\Gamma}~.
\end{equation}
The perihelion shift and the Lense-Thirring effects are defined as differences between these
orbital frequencies
\begin{equation}
\Delta_{Perihelion}=\Omega_{\phi}-\Omega_{r}=
{1\over \Gamma}\left( \sqrt{k+q^2Q^2}-{2\pi\over\omega_{r}}\right)~,
\end{equation}
\begin{equation}
\Delta_{Lense-Thirring}=\Omega_{\phi}-\Omega_{\theta}=0~.
\end{equation}
Therefore, we can see that there is no Lense-Thirring effect, due to the fact that the orbital frequencies $\Omega_{\phi}$ and $\Omega_{\theta}$ coincide  as it was found in  \cite{Grunau:2010gd, Kagramanova:2010bk}. This happens because  in the Lense-Thirring effect three rotations are involved, one of them is the rotation of the gravitating body, the other is the rotation of the test body around its own axis, and the last one is the rotation of the axis of rotation of the test body. However, in the case that we have analyzed these rotations are not present.

\section{Conclusions}
\label{summ}

In this work we studied the motion of massive particles  with electric and magnetic charges in the background of the
 the magnetically charged Garfinkle-Horowitz-Strominger
black hole.  We solved analytically the equations of motion for the test particles  in terms of the Weierstrass $\wp$, $\sigma$ and $\zeta$
elliptic functions.
We found that, the radial $r$-motion and time  $t$-motion depends on the mass and the magnetic charge of the test particle, while  angular-motion depends only on the electric charge of the test particle. We note that in the case of a test particle moving in the background of electrically charged GHS black hole, its radial motion depends on the parameters that define the test particle, that is, the mass and electric charge of the test particle \cite{Villanueva:2015kua}.

 Analyzing the effective potential, we found that bound orbits exist for $E<m$, while for $E\geq m$ the orbits are unbound, where $m$ and $E$ are the  mass and energy of the test particle.   In analogy to the Gravitational Rutherford scattering observed in  \cite{Villanueva:2015kua} we have observed the dispersion between the magnetic charge of the black hole and the magnetic charge of the test particle, which we have called Magnetic Rutherford scattering. Finally, we have studied two observables, the perihelion shift and the Lense-Thirring effects. The perihelion shift depends on electric and magnetic charges of the test particle and there is no Lense-Thirring effect due to the fact that the rotations involved in the Lense-Thirring effect are not present in the case analyzed.

The behaviour of the radial motion of the test particle except its mass, it  dependents also on two crucial parameters, the magnetic charge of the black hole Q and the magnetic charge g of the test particle. We found in the Section \ref{radial} that the stability of the radial motion of the test particle, if it crosses the horizon or if it goes to infinity  depends on a critical value of $Q_c$ and $g_c$. However, the magnetic charge Q of the GHS black hole is related to the strength of the string coupling through the relation $e^{-2\varphi}=1-\frac{Q^2}{Mr}$. Therefore, the string coupling will effect the behaviour of radial motion of the test particle.

It would be interesting to see if the magnetically charged GHS black hole can radiate its charge and what is the role of the string coupling in this effect. However, the test particle we considered in this work does not have its own dynamics to trigger such an effect. We plan to study in the lines of \cite{Gonzalez:2017shu,Kolyvaris:2017efz} the superradiance effect of a magnetically charged scalar wave scattered off the magnetically charged GHS black hole.

\begin{acknowledgments}
This work was partially funded by Comisi\'{o}n
Nacional de Ciencias y Tecnolog\'{i}a through FONDECYT Grants
11140674 (PAG) and by the Direcci\'{o}n de Investigaci\'{o}n y Desarrollo de la Universidad de La Serena (Y.V.). P. A. G. acknowledges the hospitality of the Universidad de La Serena and National Technical University of Athens and E. P. and Y. V. acknowledge the hospitality of the Universidad Diego Portales where part of this work was carried out.
\end{acknowledgments}

\appendix

\section{Hamilton-Jacobi formalism }

In this Appendix we show how the Hamilton-Jacobi formalism is connected to the Euler-Lagrange formalism.
  From the Lagrangian formulation you can construct the Hamiltonian and then perform a canonical transformation in order to obtain the Hamilton-Jacobi equation.
  In the Lagrangian formalism one define a Lagrangian from the metric
 \begin{equation}
 2 \mathcal{L} = g_{\mu \nu} \dot{x}^{\mu} \dot{x}^{\nu}=-m~,
 \end{equation}
  where $\dot{x}^{\mu}=\frac{d x^{\mu}}{d \tau}$ and $\tau$ is an affine parameter, which can be the proper time for massive particles.
The Hamiltonian is given by
\begin{equation}
\mathcal{H}=p_{\mu} \dot{x}^{\mu}-\mathcal{L}=\frac{1}{2} g^{\mu \nu} p_{\mu} p_{\nu}~, \label{haml}
\end{equation}
where $d x^{\mu} / d\tau$ was expressed in terms of the conjugate momentum and the coordinates.
Now, one can perform a canonical transformation, from the coordinates of the phase space $x^{\mu}$, $p_{\nu}$ where the Hamiltonian is $H$ to the coordinates $X^{\mu}$, $P_{\nu}$ where the Hamiltonian is $K$
\begin{equation}
p_{\mu} d x^{\mu} -H d\tau = P_{\mu} dX^{\mu} - K d\tau +dF~,
\end{equation}
where $F$ is a function of the phase space coordinates and $\tau$.
Considering the transformations
\begin{equation}
p_{\mu}=p_{\mu} (x^{\alpha},P_{\nu},\tau) \,\,\, , X^{\mu}=X^{\mu} (x^{\nu}, P_{\alpha}, \tau  )
\end{equation}
 and integrating by parts the term $P_{\mu} dX^{\mu}$, we get
 \begin{equation}
 p_{\mu} d x^{\mu} -H d\tau = - X^{\mu} dP_{\mu}- K d\tau+d (F+ X^{\mu} P_{\mu})~.\label{eq1}
 \end{equation}
 Defining the generating function $S$ of the canonical transformation as
 \begin{equation}
 S(x^{\sigma},P_{\mu}, \tau)=F+ X^{\mu} P_{\mu}~,
 \end{equation}
we obtain
 \begin{equation}
 dS=\frac{\partial S}{\partial x^{\mu}} d x^{\mu}+\frac{\partial S}{\partial P_{\mu}} dP_{\mu}+ \frac{\partial S}{\partial \tau} d \tau~.
 \end{equation}
Then equation (\ref{eq1}) gives
\begin{equation}
\left( p_{\mu}-\frac{ \partial S}{\partial x^{\mu}}  \right) d x^{\mu}-\left( \mathcal{H}+\frac{\partial S}{\partial \tau} \right) d\tau = -(X^{\mu}-\frac{\partial S}{\partial P_{\mu}}) d P_{\mu}-K d\tau~.
\end{equation}
From this equation we get
\begin{equation}
p_{\mu}=\frac{ \partial S}{\partial x^{\mu}}~,
\end{equation}
\begin{equation}
X^{\mu}=\frac{\partial S}{\partial P_{\mu}}~,
\end{equation}
\begin{equation}
K=\mathcal{H}(x^{\mu},p_{\sigma}, \tau)+\frac{\partial S}{\partial \tau}~.
\end{equation}
Setting the new Hamiltonian $K$ to zero we obtain a coordinate transformation for $S$
\begin{equation}
\mathcal{H}(x^{\mu},\frac{\partial S}{\partial x^{\sigma}}, \tau)+\frac{\partial S}{\partial \tau}=0~.
\end{equation}
Then using the Hamiltonian (\ref{haml}) we finally get
\begin{equation}
\frac{1}{2} g^{\mu \nu} \frac{\partial S}{\partial x^{\mu}} \frac{\partial S}{\partial x^{\nu}}+\frac{\partial S}{\partial \tau}=0~.\label{hamJac}
\end{equation}
In the case of the presence of electromagnetic fields the Hamilton-Jacobi equation is trivially modified.


\end{document}